# Temporal stability of intracranial EEG abnormality maps for localising epileptogenic tissue

Yujiang Wang[1,2,3,*], Gabrielle M Schroeder[1], Jonathan J Horsley[1], Mariella Panagiotopoulou[1], Fahmida A Chowdhury[3], Beate Diehl[3], John S Duncan[3], Andrew W McEvoy[3], Anna Miserocchi[3], Jane de Tisi[3], Peter N Taylor[1,2,3,*]

1. CNNP Lab (www.cnnp-lab.com), Interdisciplinary Computing and Complex BioSystems Group, School of Computing, Newcastle University, Newcastle upon Tyne, United Kingdom

2. Faculty of Medical Sciences, Newcastle University, Newcastle upon Tyne, United Kingdom

3. UCL Queen Square Institute of Neurology, Queen Square, London, United Kingdom

* *Yujiang.Wang@newcastle.ac.uk*, *Peter.Taylor@newcastle.ac.uk*

## Abstract

**Objective** Identifying abnormalities in interictal intracranial EEG, by comparing patient data to a normative map, has shown promise for the localisation of epileptogenic tissue and prediction of outcome. The approach typically uses short interictal segments of around one minute. However, the temporal stability of findings has not been established.

**Methods** Here, we generated a normative map of iEEG in non-pathological brain tissue from 249 patients. We computed regional band power abnormalities in a separate cohort of 39 patients for the duration of their monitoring period (0.92-8.62 days of iEEG data, mean 4.58 days per patient, over 4,800 hours recording). To assess the localising value of band power abnormality, we computed $D_{RS}$ - a measure of how different the surgically resected and spared tissue were in terms of band power abnormalities - over time.

**Results** In each patient, band power abnormality was relatively consistent over time. The median $D_{RS}$ of the entire recording period separated seizure free (ILAE = 1) and not seizure free (ILAE $>$ 1) patients well (AUC = 0.69). This effect was similar interictally (AUC = 0.69) and peri-ictally (AUC = 0.71).

**Significance** Our results suggest that band power abnormality $D_{RS}$, as a predictor of outcomes from epilepsy surgery, is a relatively robust metric over time. These findings add further support for abnormality mapping of neurophysiology data during presurgical evaluation.

## 1. Introduction

Interictal EEG biomarkers are currently under active research to help localise the epileptogenic zone (EZ), but a key question is their temporal stability. Increasing evidence suggests that some markers, such as EEG spikes[1] and high frequency oscillations[2], are not static, with some studies concluding that long multi-day recordings are required to observe the full range of variability in the interictal markers. To date, there has not been a direct investigation of the temporal impact on localisation ability validated with surgical outcome.

Normative band power maps have recently emerged as a promising approach to identify the EZ, and we will focus on this approach as our interictal biomarker in this study. In a given patient, the approach can be summarised as z-scoring interictally observed band power in a particular brain region to a normative control distribution of expected data in said region, thus deriving a regional band power abnormality. With intracranial EEG, obtaining a normative distribution is particularly challenging, as healthy control data are not available. Instead, the normative distribution is derived from recordings from postulated non-epileptic brain areas across a large cohort of patients[3–5]. The subsequently derived band power abnormalities have recently been shown by multiple independent studies to contain localising information[6,7,28]. However, those studies used short segments of EEG recording of around one minute temporally remote from seizures. It is unclear how the localising ability of the approach might fluctuate over time.

EEG and specifically intracranial EEG band power has been repeatedly shown to fluctuate over a range of timescales, and specifically circadian fluctuations have been consistently reported[8–12]. Previous work also highlights that EEG features may change peri-ictally[13,14]. It is therefore reasonable to assume that band power scored against a static normative map also fluctuates over time. Thus, an important next step is to investigate how band power abnormalities change over time.

In this work, we investigate band power abnormalities over time in sessions of intracranial EEG monitoring for presurgical epilepsy diagnostics, which typically span multiple days. We evaluate how temporal changes affect our ability to localise epileptogenic tissue by investigating how well we can distinguish tissue that was later resected *vs.* spared based on band power abnormality in each brain region over time. We further test if there are specific peri-ictal changes in our ability to distinguish resected and spared tissue and validate all results with patient outcomes of post-surgical seizure freedom.

## 2. Methods

Our approach is to compute a normative map of interictal intracranial EEG band power to which patients from an independent site can be compared. Then, in comparing patients to the normative map we can compute band power abnormality, with the expectation that if abnormalities are present beyond the resection the patient will not be seizure-free. We use pre-operative MRI, post-implant CT, and post-operative MRI to localise electrodes to parcellated brain regions and identify resection margins. We compare resected and spared band power abnormalities using the $D_{RS}$ statistic, and investigate its consistency over time (up to nine days). A summary of the processing steps is shown in figure *1*.

### 2.1 Patients

We analyzed iEEG data from two cohorts: 39 patients with refractory focal epilepsy from the University College London Hospital (UCLH) (Table *1*), and 249 patients from the Restoring Active Memory (RAM) data set. Our processing is broadly similar to previous work[6]. We created a normative map of iEEG band power from the RAM iEEG data. The normative map was used as a baseline to compute the time-varying iEEG band power abnormalities of the UCLH patients. Data were analysed following approval from the Newcastle University Ethics Committee (2225/2017).

**Table 1: Summary of UCLH patient data.**

| | ILAE1 | ILAE>1 | Test statistic |
|---|---|---|---|
| **N (%)** | 16(41%) | 23(59%) | |
| **Age (mean,SD)** | 29.7 (4.3) | 31.8 (9.5) | p=0.41, t=-0.84 |
| **Sex (M,F)** | 7,9 | 11,12 | p=0.80, $\chi^2$=0.06 |
| **Temporal, extratemporal** | 8,8 | 11,12 | p=0.89, $\chi^2$=0.02 |
| **Side (Left, Right)** | 10,6 | 13,10 | p=0.71, $\chi^2$=0.14 |
| **Num contacts (mean, sd)** | 76.9 (27.5) | 63.3 (22.8) | p=0.1, t=1.68 |
| **Recording Duration in hours (mean, sd)** | 122.9 (56.1) | 123.1 (43.9) | p=0.99, t=-0.01 |

### 2.2 MRI processing

To generate a normative map we localised RAM electrode coordinates to regions as described previously[6]. In brief, we used FreeSurfer to generate volumetric parcellations of an MNI space template brain[15,16]. Each electrode contact was assigned to the closest grey matter volumetric region within 5 mm. If the closest grey matter region was >5mm away then the contact was excluded from further analysis. For UCLH data a similar technique was used but applied in native space using the patient's own parcellated pre-operative MRI.

To identify which regions were later resected in the UCLH cohort we used previously described methods[6,17]. We registered post-operative MRI to the pre-operative MRI and manually delineated the resection cavity. This manual delineation accounted for post-operative brain shift and sagging into the resection cavity. Electrode contacts within 5mm of the resection were assigned as resected. Regions with >25% of their electrode contacts removed were considered as resected for downstream analysis.

## 2.3 iEEG processing

From each RAM patient, we extracted a single 30 sec segment of interictal iEEG data, recorded during a period of relaxed wakefulness. To approximate non-pathological brain dynamics, we excluded electrodes located in lesions or the seizure onset zone. We additionally removed visually and algorithmically identified noisy electrodes from the analysis. Each segment was then re-referenced to a common average reference, notch filtered at 60 Hz (2 Hz width, fourth order zero-phase Butterworth filter), band pass filtered from 0.5 to 80 Hz (fourth order zero-phase Butterworth filter), and downsampled to 200 Hz.

In the UCLH cohort, we first divided each patient's continuous iEEG data into 30 s non-overlapping, consecutive time windows. We referenced each window of data to a common average reference, with any noisy channels (with outlier amplitude ranges) excluded from the computed average. Each segment was then notch filtered at 50 Hz, band pass filtered from 0.5 to 80 Hz (fourth order zero-phase Butterworth filter), and downsampled to 200 Hz. Time windows with missing data were omitted from the analysis.

We then computed the iEEG band power of both the RAM and UCLH iEEG data. We first computed the power spectral density (PSD) of each electrode contact in each 30 s iEEG segment with 2 s, non-overlapping windows. From each PSD, we used Simpson's rule to compute band power in five frequency bands: delta (1-4 Hz), theta (4-8 Hz), alpha (8-13 Hz), beta (13-30 Hz), gamma (30-47.5 Hz, 52.5-57.5 Hz, 62.5-77.5 Hz). We chose the gamma band limits to omit electrical noise frequencies. We $\log_{10}$ transformed the band power values and, for each iEEG segment and channel, normalized the set of five band power values to sum to one, producing the relative log band power for each frequency band.

### 2.3.1 Creating iEEG band power normative map

To produce a normative map of relative log band power values, we averaged this measure across electrodes and patients within each ROI. First, for each RAM patient, we took the mean relative log band power across all of the patient's electrodes within each ROI. This step yield patient-specific relative log band power values at the ROI, rather than electrode, level. The normative map was then defined by the mean $\mu_{f,i}$ and standard deviation $\sigma_{f,i}$ of the relative log band power in frequency band $f$ and ROI $i$ across the RAM patients. Regions with coverage from fewer than five subjects were excluded from the normative map and further analysis.

### 2.3.2 Computing time-varying abnormalities and $D_{RS}$

As with the RAM patients, we first computed relative log band power values at the ROI level for each UCLH patient within each 30 s time window. As before, this transformation was achieved by taking the mean relative log band power across all of the patient's electrodes in ROI $i$. For each frequency band $f$, ROI $i$, and time window $t$, we then computed a $z$-score $z_{f,i,t}$ by standardising the patient's relative log band power $b_{f,i,t}$ by the normative map:

$$z_{f,i,t} = \frac{b_{f,i,t} - \mu_{f,i}}{\sigma_{f,i}}$$

We then defined the patient's band power abnormality for each ROI and time window as the maximum absolute z-score across the five frequency bands. Thus, each UCLH patient's iEEG recording was described by time-varying abnormalities in their ROIs with electrode coverage.

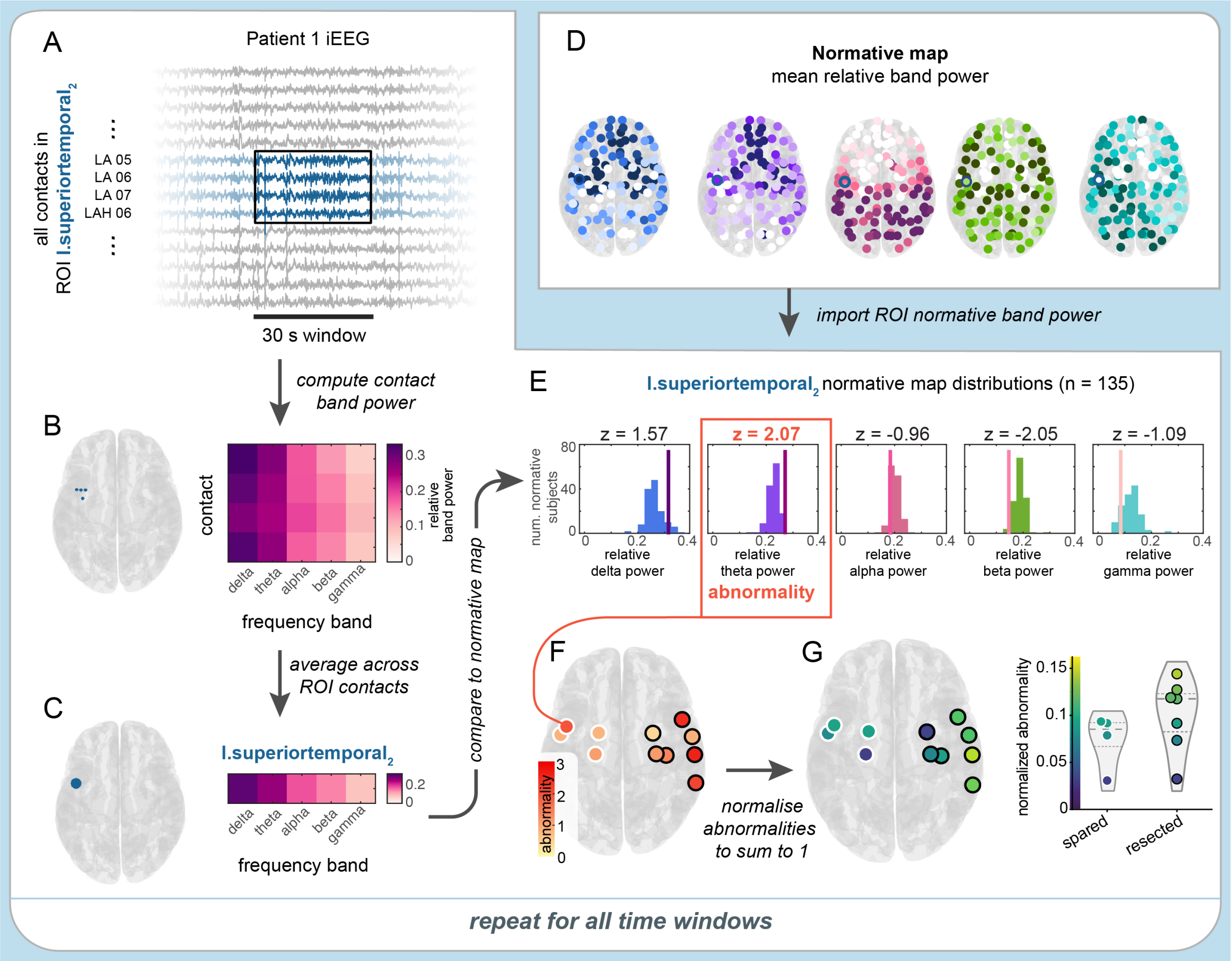

***Figure 1: Computing band power abnormality in an example time window and ROI in Patient 1.*** *A) Example 30 s time window of iEEG data in a subset of Patient 1's contacts. All contacts within an example ROI, l.superiortemporal2, are highlighted in blue. B) From the 30 s of iEEG data, the relative log band power of each of the four contacts in the example ROI was computed. C) Averaging relative log band power across all of the ROI's contacts produces the ROI's relative band power. D) Relative log band power was also computed in a separate cohort of 249 subjects, yielding a normative map of this measure. E) Patient 1's ROI relative band power was then z-scored relative to the normative map.The ROI's abnormality was defined as the maximum absolute z-score (here, 2.07) across the five frequency bands. F) The process is repeated for all ROIs. G) The abnormality values are normalized so their sum equals 1 and plotted for resected and spared regions. This process was repeated for all time windows in each patient.*

To quantify the level of band power abnormalities in resected and spared ROIs, we computed the Distinguishability of Resected and Spared tissue ($D_{RS}$) for each time window in each UCLH patient. We defined $D_{RS}$ as the area under the curve (AUC) for distinguishing resected and spared ROIs using band power abnormalities, with $D_{RS}$ <0.5 indicating higher abnormalities in resected ROIs and $D_{RS}$ >0.5 revealing higher abnormalities in spared ROIs.

#### 2.3.3 Identifying interictal and peri-ictal periods

For each UCLH patient, we labelled each time window as ictal, interictal, or peri-ictal based on the patient's seizure times. From each patient's clinical annotations and reports, we obtained the times and durations of all recorded seizures, including subclinical seizures. The 30 s time windows containing seizures were labelled ictal windows. Time windows within one hour of an ictal time window, excluding the ictal time windows themselves, were labelled peri-ictal windows. Finally, the remaining time windows were labelled interictal windows.

## 3. Results

We analyzed relative band power abnormalities in continuous iEEG data of 39 patients with focal epilepsy who underwent surgical resection. We focused on the presence of abnormalities in spared and resected brain regions in each patient, as captured by our measure $D_{RS}$. In the following sections, we first present time-varying abnormalities and $D_{RS}$ in two example patients. We then show the level of $D_{RS}$ variability across patients and relate typical $D_{RS}$ values to patient surgical outcomes. Finally, we compare $D_{RS}$ values in interictal and peri-ictal periods.

### 3.1 Location of abnormalities remain relatively stable

Fig. 2A shows the time-varying abnormalities of an example patient (1), who was seizure free following surgery (ILAE = 1). While there was some spatial variability in abnormalities across the recording, abnormalities tended to be higher in resected brain regions. As such, $D_{RS}$ was below 0.5 in most time windows, and the median $D_{RS}$ across the recording was 0.29 (Fig. 2B). Fig. 2C and D, which show the abnormalities of an example time window with $D_{RS} = 0.29$, further demonstrate the presence of higher abnormalities in resected brain regions in this patient.

Meanwhile, patient 2 had a poor surgical outcome (ILAE = 4). As in patient 1, regional abnormalities were relatively consistent across patient 2's recording (Fig. 2E). However, in patient 2, higher abnormalities were located in spared regions, producing a $D_{RS}$ above 0.5 across almost the entire recording and a high median $D_{RS}$ of 0.8 (Fig. 2F-H). Thus, in both patients, the median $D_{RS}$ corresponds to post-surgical outcome and is relatively stable over time.

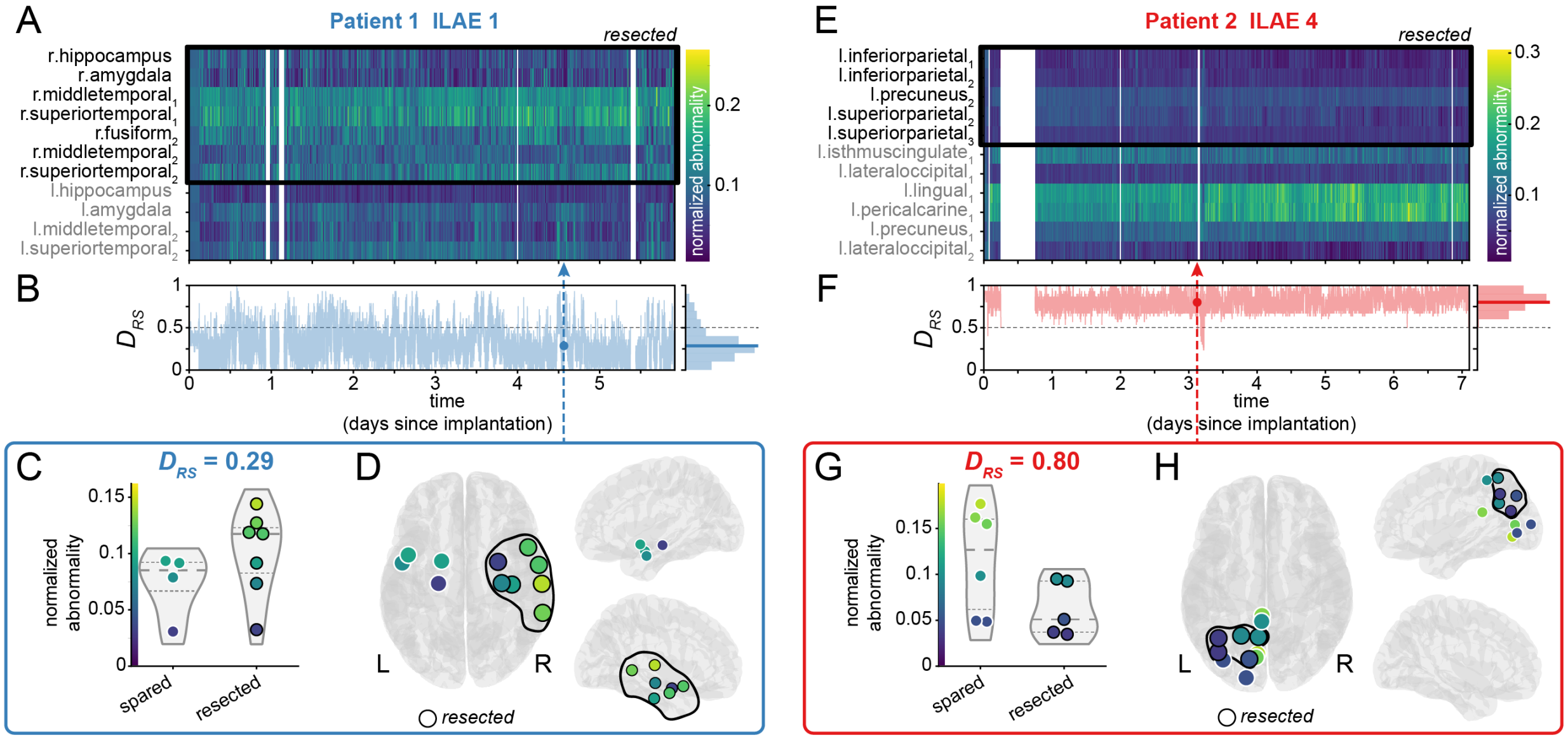

***Figure 2: Time-varying*** $D_{RS}$ ***in example patients.*** *Time-varying abnormalities and* $D_{RS}$ *in example patients 1 (A-D) and 2 (E-H). A and E) Heatmap of regional maximum absolute band power abnormalities, with each column corresponding to a 30 s time window in the patient's iEEG recording. Abnormalities in each time window are normalized to sum to one, thus showing each region of interest's (ROI's) contribution to the total abnormality in that time window. Resected ROIs are outlined with a black box. B and F) Time-varying* $D_{RS}$ *computed from band power abnormalities. Histogram to the right of each plot shows the distribution of* $D_{RS}$ *in each recording, with the median* $D_{RS}$ *marked with a bold horizontal line. The circle and dashed vertical line mark an example time window that had a* $D_{RS}$ *equal to the patient's median* $D_{RS}$*.* $D_{RS} = 0.5$ *is also shown with a dashed black line for reference. C and G) Normalized abnormalities of spared and resected ROIs in the example time window with* $D_{RS}$ *equal to the patient's median* $D_{RS}$*. Quartiles of the abnormality distributions are marked with dashed lines. (D) and (H) show the same abnormalities on a brain surface from top and side views.*

## 3.2 Median $D_{RS}$ over time separates surgical outcomes

We next investigated variability in $D_{RS}$ in all 39 patients in our cohort. Fig. 3A shows the distribution of $D_{RS}$ values in each patient's iEEG recording. While there was within-patient variability in $D_{RS}$, most distributions were unimodal, with $D_{RS}$ fluctuating around a particular value. Thus, although $D_{RS}$ could differ between time windows, it was not random or highly variable.

As a simple measure of consistency in $D_{RS}$, we computed the percentage of time windows in each patient with a $D_{RS}$ value less than or equal to 0.5 (Fig. 3B). We refer to this value as the “localizing percentage" of time windows; a high percentage indicates that abnormalities were consistently higher in the patient’s resected brain regions, and thus localized to the hypothesized epileptogenic zone. For example, in patient 1 (Fig. 2A-D), the localizing percentage was 87.2% due to consistently high abnormalities in resected regions. Low localizing percentages, which resulted from consistently high $D_{RS}$ values, revealed that abnormalities were instead usually higher in spared regions, as in patient 2 (Fig. 2E-H, localizing percentage = 0.3%). In 28 of the 39 patients (72%), the localizing proportion was either low ($\leq$ 20%) or high ($\geq$ 80%). Thus, in most patients, $D_{RS}$ values were either consistently above or below 0.5, indicating a consistent relationship between spared and resected abnormalities.

We then determined whether each patient’s typical $D_{RS}$ value, as captured by the median of their $D_{RS}$ distribution, was associated with surgical outcome. Median $D_{RS}$ was higher in patients who were not seizure free (ILAE 2-5) versus seizure free (ILAE 1) after surgical resection ($p$ = 0.021, one-sided Wilcoxon rank sum test) (Fig. 3C), and the area under the curve (AUC) when using $D_{RS}$ as a binary classifier of patient surgical outcome was 0.69 (Fig. 3D). Patients who were not seizure free had median $D_{RS}$ values higher than 0.5 ($p$ = 0.006, one-sided Wilcoxon signed rank test), indicating higher abnormalities in spared brain regions in this group. However, seizure free patients did not have median $D_{RS}$ values lower than 0.5 ($p$ = 0.455, one-sided Wilcoxon signed rank test).

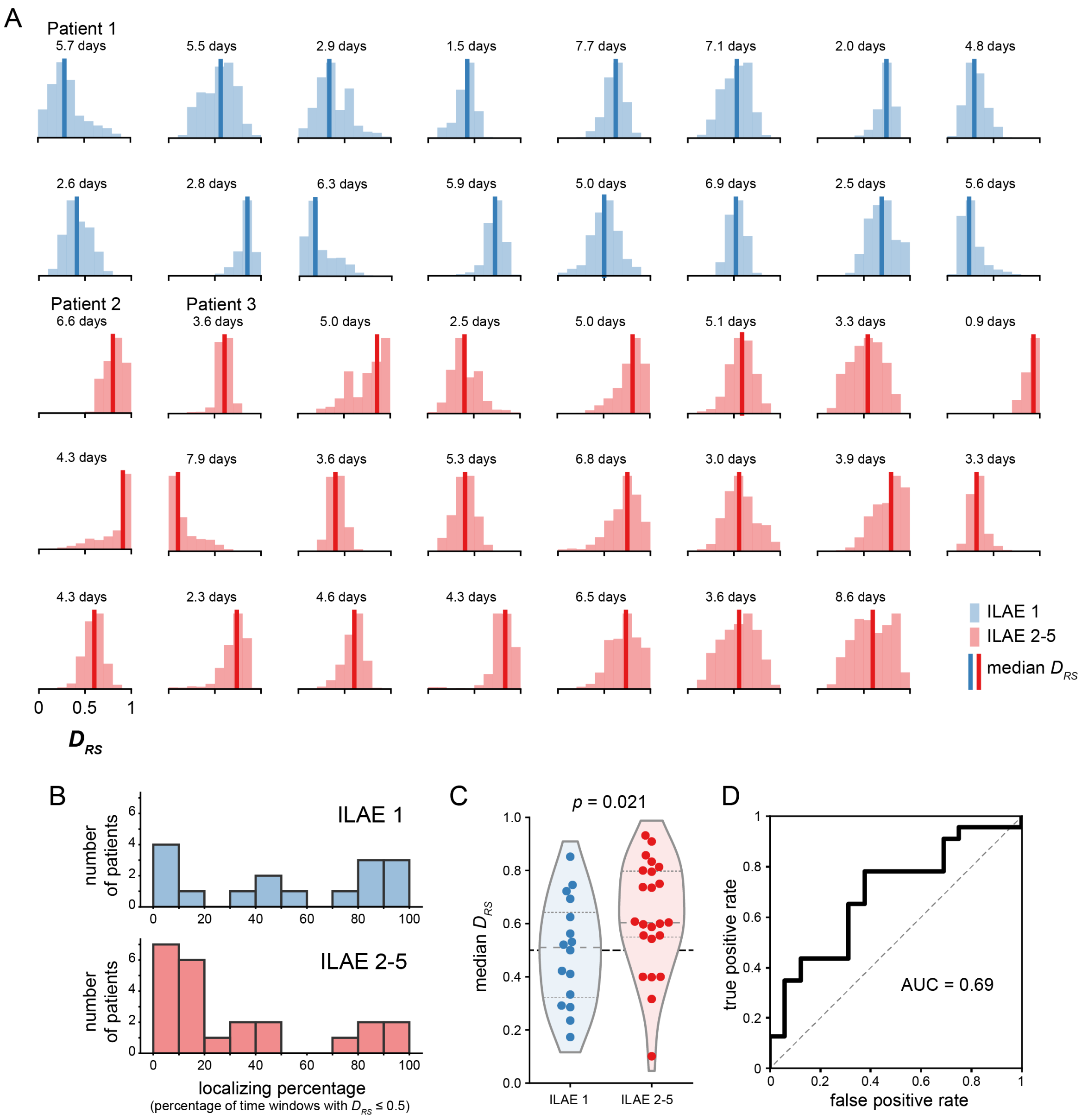

***Figure 3: Variability in*** $D_{RS}$ ***across patients.*** *A) Distribution of* $D_{RS}$ *in each patient's iEEG recording, as shown in example patients in Fig. 2B and F. Bold vertical lines show the median of each distribution. Number of days of data used to compute each distribution is also provided. B) Localizing percentage of time windows (i.e., percentage of time windows with* $D_{RS} \leq 0.5$*) in our cohort. C) Comparison of median* $D_{RS}$ *in patients who were seizure free (ILAE 1) versus not seizure free (ILAE 2-5) after surgery. Quartiles of the* $D_{RS}$ *distributions are marked with dashed lines. D) Receiver operator characteristic (ROC) curve using median* $D_{RS}$ *as a binary classifier of patient surgical outcome.*

### 3.3 Interictal and peri-ictal time windows perform similarly at distinguishing patient surgical outcomes

Finally, we determined whether $D_{RS}$ differed between interictal and peri-ictal (defined as within one hour of a seizure) periods within each patient. Fig. 4A shows the time-varying abnormalities of patient 3, with seizure times marked with red dashed lines. The pattern of abnormalities appears relatively similar across the recording, regardless of the proximity to seizures. Likewise, patient 3's $D_{RS}$ was similar in interictal and peri-ictal periods (Fig. 4B), and the patient had almost the same median interictal and peri-ictal $D_{RS}$ (0.61 and 0.60, respectively).

Across patients, we also observed that each patient had similar median interictal and median peri-ictal $D_{RS}$ (Fig. 4C). The median values of each time period were not different either across all patients or within patients with the same surgical outcome ($p$ = 0.17 for all patients, $p$ = 0.09 for ILAE 1 patients, $p$ = 0.68 for ILAE 2-5 patients, two-sided Wilcoxon signed rank tests). As such, median interictal and median peri-ictal $D_{RS}$ also performed similarly at distinguishing patients by their surgical outcomes with AUCs of 0.69 and 0.71, respectively ($p$ = 0.022 and $p$ = 0.014, one-sided Wilcoxon rank sum tests) (Fig. 4D).

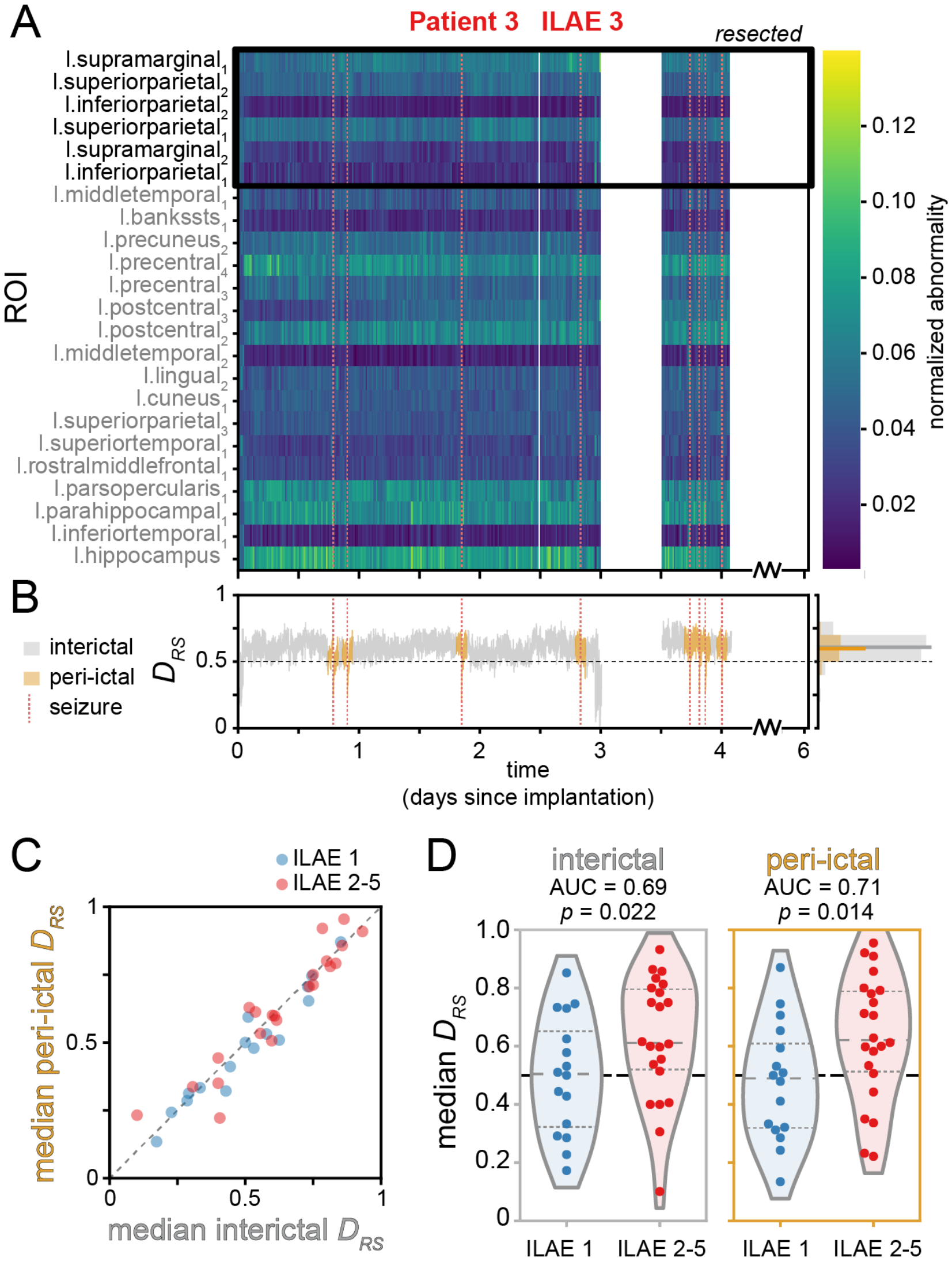

***Figure 4:*** ***$D_{RS}$* in interictal and peri-ictal periods.** *A-B) Interictal and peri-ictal $D_{RS}$ in example patient 3. Seizure times are marked with vertical dashed red lines. A) Heatmap of time-varying regional band power abnormalities (normalized to sum to one). Resected ROIs are outlined with a black box. B) Time-varying $D_{RS}$, colored by whether the time window was interictal (gray) or peri-ictal (orange). Histogram to the right shows the distribution of $D_{RS}$ and median value (bold horizontal lines) in each time period. C-D) Interictal and peri-ictal $D_{RS}$ across patients. C) Median peri-ictal $D_{RS}$ versus median interictal $D_{RS}$ of each patient, colored by patient surgical outcome. D) Comparison of median interictal (left) and peri-ictal (right) $D_{RS}$ in patients who were seizure free versus not seizure free after surgery. Quartiles of the $D_{RS}$ distributions are marked with dashed lines.*

## 4. Discussion

We have investigated the temporal stability of interictal band power abnormality patterns on intracranial EEG relative to a normative map over timescales of multiple days. We demonstrated that the spatial distribution of abnormalities is temporally stable in terms of its localisation ability, even peri-ictally. Furthermore, we reproduced the previously reported separation of patients by post-surgical outcome based on median distinguishability of spared and resected tissue in each patient.

These findings have important implications for the practical application of interictal abnormality detection in pre-surgical evaluation with iEEG. Our results suggest that estimating a median abnormality map is sufficent to obtain localising information. This could be achieved with, for example, randomly sampling short segments of around one minute of icEEG. Ideally several random segments are used, spaced out in time. Interestingly, even peri-ictal segments could be used to obtain an interictal abnormality map, although we still suggest avoiding segments with obvious ictal and peri-ictal phenomena. Practically speaking, the suitability of peri-ictal segments makes adopting the method immediately feasible, as it does not require additional manual screening or large amounts of continuous icEEG data handling. Beyond invasive iEEG recordings, our results hint at the possibility that noninvasive abnormality maps are likely also relatively stable over time. This hypothesis is supported by a recent MEG study mapping abnormalities across multiple epochs[28].

Our findings are encouraging, especially when compared to other traditional interictal markers of epileptogenic tissue such as interictal spikes. Apart from the previously-demonstrated added value of band power abnormality compared to interictal spikes in our cohort[6], our results here also seem to suggest that band power abnormality is temporally stable enough as a biomarker. It is unknown if a similar stability is seen with interictal spikes, as spike load is reported to be generally higher during sleep[18], the pattern and location can vary depending on brain state[1], and a minimum analysis period of at least 24h has been recommended[1]. We encourage future work to evaluate and compare the temporal stability of more electrophysiological markers directly in the context of localisation and validate it with surgical resection and outcome information.

Despite the temporal stability in terms of localisation ability, our data also clearly shows some level of temporal fluctuations in band power abnormalities. These fluctuations do not hamper the relative stability of distinguishability of resected and spared tissue. Nevertheless, temporal fluctuations are present, and in some examples clearly structured in time (see e.g. Fig. *2*A, region l.middletemporal2 for a circadian fluctuation in band power abnormality). These temporal fluctuations could reveal different pathological subnetworks and relate to seizure occurrence[19], for example when they coalesce in space and time[20–22]. It is also possible that they are related to seizure severity or other temporally modulated properties of seizures[8,23,24]. The notion of fluctuating pathological subnetworks may also suggest alternative treatment strategies of network resections, disconnections or closed-loop neuromodulation of subnetworks[20–22]. Finally, these abnormalities may not be epileptogenic *per se*, but relate to other temporally changing impairments in e.g. mood or cognitive performance. Future research should not neglect these time-varying aspects simply due to our reported temporal stability in terms of localisation performance, as these aspects may be the key to understanding fundamental mechanisms of epilepsy.

An important step for making band power abnormalities more specific and predictive is to account for the normative map of band power in various brain states. Sleep and vigilance states must be accounted for, as their band power changes are well-known and described in healthy humans[25]. Cognitive and mood states have also been reported in terms of electrographic correlates[26,27]. We expect that by accounting for brain states, and other potential confounds, the abnormalities will become more specific to epileptogenic tissue. In clinical practice, we envisage one or multiple short, well-controlled abnormality mapping paradigms during icEEG monitoring with a straightforward, but well-defined state/task.

In summary, using continously-recorded icEEG, we have demonstrated that band power abnormality maps are temporally stable in terms of their localising information, despite brain state and seizure-related changes throughout the recording. This finding is an important cornerstone in establishing feasibility of band power abnormality mapping to aid localisation for presurgical evaluation. We encourage investigating temporal patterns in band power for increased predictive power and mechanistic insights into epilepsy.

## 5. Acknowledgements

We thank members of the Computational Neurology, Neuroscience & Psychiatry Lab (www.cnnp-lab.com) for discussions on the analysis and manuscript; P.N.T. and Y.W. are both supported by UKRI Future Leaders Fellowships (MR/T04294X/1, MR/V026569/1). MP and JJH are supported by the Centre for Doctoral Training in Cloud Computing for Big Data (EP/L015358/1). JSD, JdT are supported by the NIHR UCLH/UCL Biomedical Research Centre. B.D. receives support from the NIH National Institute of Neurological Disorders and Stroke U01-NS090407 (Center for SUDEP Research) and Epilepsy Research UK.

## 7. References

1. Conrad EC, Tomlinson SB, Wong JN, Oechsel KF, Shinohara RT, Litt B, et al. Spatial distribution of interictal spikes fluctuates over time and localizes seizure onset. Brain. 2020; 143(2):554–69.

2. Gliske SV, Irwin ZT, Chestek C, Hegeman GL, Brinkmann B, Sagher O, et al. Variability in the location of high frequency oscillations during prolonged intracranial EEG recordings. Nature communications. 2018; 9(1):1–4.

3. Betzel RF, Medaglia JD, Kahn AE, Soffer J, Schonhaut DR, Bassett DS. Structural, geometric and genetic factors predict interregional brain connectivity patterns probed by electrocorticography. Nature biomedical engineering. 2019; 3(11):902–16.

4. Frauscher B, Von Ellenrieder N, Zelmann R, Doležalová I, Minotti L, Olivier A, et al. Atlas of the normal intracranial electroencephalogram: Neurophysiological awake activity in different cortical areas. Brain. 2018; 141(4):1130–44.

5. Groppe DM, Bickel S, Keller CJ, Jain SK, Hwang ST, Harden C, et al. Dominant frequencies of resting human brain activity as measured by the electrocorticogram. Neuroimage. 2013; 79:223–33.

6. Taylor PN, Papasavvas CA, Owen TW, Schroeder GM, Hutchings FE, Chowdhury FA, et al. Normative brain mapping of interictal intracranial EEG to localize epileptogenic tissue. Brain. 2022; 145(3):939–49.

7. Bernabei JM, Sinha N, Arnold TC, Conrad E, Ong I, Pattnaik AR, et al. Normative intracranial EEG maps epileptogenic tissues in focal epilepsy. Brain.

8. Panagiotopoulou M, Papasavvas C, Schroeder GM, Taylor P, Wang Y. Fluctuations in EEG band power over minutes to days explain how seizures change over time. arXiv:201207105 [q-bio] [Internet]. 2020 [cited 2021]; Available from: *http://arxiv.org/abs/2012.07105*

9. Aeschbach D, Matthews JR, Postolache TT, Jackson MA, Giesen HA, Wehr TA. Two circadian rhythms in the human electroencephalogram during wakefulness. American Journal of Physiology-Regulatory, Integrative and Comparative Physiology. 1999; 277(6):R1771–9.

10. Geier C, Lehnertz K. Long-term variability of importance of brain regions in evolving epileptic brain networks. Chaos: An Interdisciplinary Journal of Nonlinear Science. 2017; 27(4):043112.

11. Geier C, Lehnertz K, Bialonski S. Time-dependent degree-degree correlations in epileptic brain networks: From assortative to dissortative mixing. Frontiers in Human Neuroscience. 2015; 9:462.

12. Mitsis GD, Anastasiadou MN, Christodoulakis M, Papathanasiou ES, Papacostas SS, Hadjipapas A. Functional brain networks of patients with epilepsy exhibit pronounced multiscale periodicities, which correlate with seizure onset. Human brain mapping. 2020; 41(8):2059–76.

13. Pearce A, Wulsin D, Blanco JA, Krieger A, Litt B, Stacey WC. Temporal changes of neocortical high-frequency oscillations in epilepsy. Journal of neurophysiology. 2013; 110(5):1167–79.

14. Perucca P, Dubeau F, Gotman J. Widespread EEG changes precede focal seizures. PloS one. 2013; 8(11):e80972.

15. Fischl B. FreeSurfer. Neuroimage. 2012; 62(2):774–81.

16. Hagmann P, Cammoun L, Gigandet X, Meuli R, Honey CJ, Wedeen VJ, et al. Mapping the structural core of human cerebral cortex. PLoS biology. 2008; 6(7):e159.

17. Taylor PN, Sinha N, Wang Y, Vos SB, Tisi J de, Miserocchi A, et al. The impact of epilepsy surgery on the structural connectome and its relation to outcome. NeuroImage: Clinical. 2018; 18:202–14.

18. Ng M, Pavlova M. Why are seizures rare in rapid eye movement sleep? Review of the frequency of seizures in different sleep stages. Epilepsy research and treatment. 2013; 2013.

19. Karoly PJ, Rao VR, Gregg NM, Worrell GA, Bernard C, Cook MJ, et al. Cycles in epilepsy. Nature Reviews Neurology. 2021; 17(5):267–84.

20. Spencer SS. Neural networks in human epilepsy: Evidence of and implications for treatment. Epilepsia. 2002; 43(3):219–27.

21. Wang Y, Goodfellow M, Taylor PN, Baier G. Dynamic mechanisms of neocortical focal seizure onset. PLoS computational biology. 2014; 10(8):e1003787.

22. Wang Y, Trevelyan AJ, Valentin A, Alarcon G, Taylor PN, Kaiser M. Mechanisms underlying different onset patterns of focal seizures. PLoS computational biology. 2017; 13(5):e1005475.

23. Schroeder GM, Diehl B, Chowdhury FA, Duncan JS, De Tisi J, Trevelyan AJ, et al. Seizure pathways change on circadian and slower timescales in individual patients with focal epilepsy. Proceedings of the National Academy of Sciences. 2020; 117(20):11048–58.

24. Gascoigne SJ, Waldmann L, Panagiotopoulou M, Chowdhury F, Cronie A, Diehl B, et al. A library of quantitative markers of seizure severity. arXiv preprint arXiv:220615283. 2022;

25. Roth B. The clinical and theoretical importance of EEG rhythms corresponding to states of lowered vigilance. Electroencephalography & Clinical Neurophysiology. 1961;

26. Wang S, Zhu Y, Wu G, Ji Q. Hybrid video emotional tagging using users' EEG and video content. Multimedia tools and applications. 2014; 72(2):1257–83.

27. Hu X, Yu J, Song M, Yu C, Wang F, Sun P, et al. EEG correlates of ten positive emotions. Frontiers in human neuroscience. 2017; 11:26.

28. Owen, T.W., Schroeder, G.M., Janiukstyte, V., Hall, G.R., McEvoy, A., Miserocchi, A., de Tisi, J., Duncan, J.S., Rugg-Gunn, F., Wang, Y., Taylor, P.N.,. MEG abnormalities highlight mechanisms of surgical failure in neocortical epilepsy. Epilepsia [in press]. 2023